\documentclass[%
 aip,
 amsmath,amssymb,
 reprint,%
]{revtex4-2}
\usepackage{epsfig,amssymb,amsfonts}
\usepackage{amsmath,amsfonts,amssymb}
\usepackage{subfigure}
\usepackage{graphicx}
\usepackage{xcolor}
\input{epsf}
\usepackage{epstopdf}

\begin{document}

\title{Coherent preparation of the biexciton state in a semiconductor quantum dot coupled to a metallic nanoparticle}

\author{Emmanuel Paspalakis}
\email{paspalak@upatras.gr}

\author{Athanasios Smponias}

\author{Dionisis Stefanatos}
\email{dionisis@post.harvard.edu}

\affiliation{Materials Science Department, School of Natural Sciences, University of Patras, Patras 26504, Greece}

\date{\today}

\begin{abstract}
We study the potential for controlled transfer of population to the biexciton state of a semiconductor quantum dot coupled with a metal nanoparticle, under the influence of an electromagnetic pulse with hyperbolic secant shape, and derive analytical solutions of the density matrix equations, for both zero and nonzero biexciton energy shift. These solutions lead to efficient transfer to the biexciton state, for various interparticle distances, including relatively small values. In certain cases, when the distance between the two particles is small, the transfer of population is strongly modified because of the influence of surface plasmons to the excitons, and the effect is more pronounced for shorter pulses. The hybrid snanostructure that we study has been proposed for generating efficiently polarization-entangled photons, thus the successful biexciton state preparation considered here is expected to contribute in this line of research.
\end{abstract}



\maketitle
\bibliographystyle{apsrev}

\section{Introduction}

In the area of quantum plasmonics \cite{Tame13a,Review}, emphasis has been given to the controlled dynamics of the population of states of semiconductor quantum dots which interact with electromagnetic fields, while they are also coupled with metal nanoparticles \cite{Cheng07a,Sadeghi09a,Sadeghi10a,Malyshev13a,Carreno18a,Anton12a,Anton13a,Paspalakis13a,Lee15a,McMillan16a,Qi19a,Smbonias21a}. It has been found that the population dynamics for a quantum dot next to a metallic nanoparticle is markedly different from the population dynamics of an isolated quantum dot. The reason is the strengthening of electric field as well as the strong interaction between excitons and surface plasmons, which take place when metal nanoparticles are present \cite{Wang06a,Yan08a,Miri18a,Artuso10a,Malyshev11a,Hatef12a,Kosionis13a,Zhao14a,Singh16a,Kosionis2019,You2019,Singh20a}. The main problem of interest in the controlled population dynamics deals with the application of external fields to the ground and single exciton energy levels of the quantum dot \cite{Cheng07a,Sadeghi09a,Sadeghi10a,Malyshev13a,Carreno18a,Anton12a,Paspalakis13a,Lee15a,McMillan16a,Smbonias21a}, giving emphasis to the influence of the nanoparticle to the transfer of population between these two levels. The period of the corresponding Rabi cycle is significantly altered due to the presence of a metal nanoparticle \cite{Cheng07a,Sadeghi09a,Sadeghi10a,Malyshev13a,Carreno18a}, and plasmonic meta-resonances may even destroy Rabi oscillations for certain interparticle distances \cite{Sadeghi09a,Malyshev13a,Carreno18a}. Also, high efficiency exciton transfer in coupled quantum dot-metal nanoparticles can be created for specific short \cite{Paspalakis13a,McMillan16a,Smbonias21a} and ultra-short \cite{Lee15a,McMillan16a,Smbonias21a} applied electromagnetic pulses. In addition, electromagnetically-induced selective excitonic population transfer can be achieved in a quantum dot when a metal nanoparticle is present \cite{Anton12a}.
Optimal control has been used to maximize the transfer of population among the lower energy levels in a $\Lambda$-type quantum dot, placed close to a metallic nanoparticle \cite{Qi19a}. Furthermore, the combination of pulsed and continuous wave fields has been proposed for high efficiency preparation of a single hole spin state in a semiconductor quantum dot near a metal nanoparticle \cite{Anton13a}.

Another important problem in the area of semiconductor quantum dots is the coherent generation of the biexciton state when starting from the ground state \cite{Flissikowki04a,Akimov06a,Stufler06a,Hui08a,Machnikowski08a,Paspalakis10a,Axt13a,Amand13a,Brumer13a,Axt14a,Ardelt14,Quilter15,Axt15a,Kaldewey17a,Stefanatos20a}. A simple method is to apply one laser pulse with linear polarization, implementing a two-photon transition from the ground state to the biexciton state. This approach can be useful for generating efficiently single photons \cite{Weihs13a} and polarization-entangled photons \cite{Muller14a,Heinze15a,Winik17a,Huber17a,Chen18a} for quantum information processing.

Although the exciton-biexciton cascade in quantum dots next to metal nanoparticles, which act as plasmonic nanoantennas, has been proposed for the efficient generation of polarization-entangled photons \cite{Maksymov12a} and the strong enhancement of biexciton emission \cite{Sandoghdar16a,Krivenkov19a}, the studies on coherent preparation of biexciton state and on controlled biexciton dynamics in a quantum dot, placed close to a metallic nanoparticle, are limited. To the best of our knowledge, this problem is studied only in the work of Nughoro {\it et al.} \cite{Nugroho19a}, where the two-photon Rabi oscillations have been analyzed.
The purpose of the present work is to study the coherent preparation of the biexciton state in a small semiconductor quantum dot which is placed close to a spherical metallic nanoparticle. We model the quantum dot as a three-level ladder-type quantum system, whose interaction with the applied electromagnetic pulses is described using the modified nonlinear density matrix equations, which incorporate the interaction between surface plasmons on the nanoparticle and excitons in the quantum dot \cite{Wang06a,Yan08a,Artuso10a,Paspalakis13a,Nugroho19a,Malyshev15a}. The applied electromagnetic field is linearly polarized, at two-photon resonance with the ground-biexciton transition, and the pulses have hyperbolic secant envelope. For these control pulses, we present here analytical solutions of the density matrix equations for both zero and nonzero biexciton energy shift, which efficiently populate the biexciton level for various distances separating the metallic nanoparticle and the semiconductor quantum dot. The method is successful even for small interparticle distances. We point out that the coherent control of quantum dots and transmon qubits which interact with hyperbolic secant pulses has been studied both theoretically \cite{Hui08a,Paspalakis10a,Paspalakis04b,Kis04a,Economou06a,Economou20a} and experimentally \cite{Economou09a,Poem11a,Pappas17a}. In addition, electromagnetic pulses with hyperbolic secant shape have been exploited for the efficient transfer of population from the ground to the exciton state in a semiconductor quantum dot, with \cite{Paspalakis13a} and without \cite{Paspalakis06a} the presence of a metallic nanoparticle.

The article has the following structure: in section II we introduce the density matrix equations describing the interaction of the hybrid nanostructure with the applied pulsed field. Subsequently, in section III, we discuss results which are derived by solving analytically and numerically the density matrix equations. We highlight examples where high levels of transfer fidelity to the biexciton state are obtained. In section IV we summarize our findings. We also give an Appendix, where details for the analytical solutions shown in the manuscript are presented.

\section{Theory of the hybrid nanostructure}

\begin{figure}[t]
\centering
\includegraphics[width=3in]{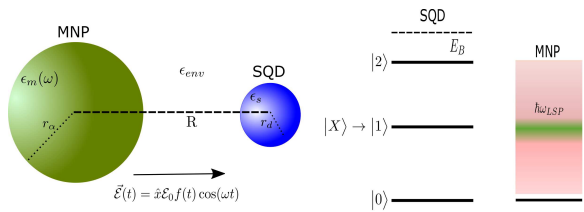}
\caption{Left: The nanosystem that we study consists of a spherical metallic nanoparticle and a spherical semiconductor quantum dot with radii $r_a, r_d$ and dielectric constants $\varepsilon_m(\omega), \varepsilon_s$, respectively. The distance between the centers of the particles is $R$. The dielectric constant of the surrounding environment is denoted by $\varepsilon_{env}$, while the applied field is $\mathcal{\vec{E}}(t)$. Right: Energy-level diagram of the system.}
\label{fig1}
\end{figure}

Fig.\ \ref{fig1} displays the nanostructure under consideration, consisting of a spherical metal nanoparticle and a spherical semiconductor quantum dot of radii $r_{a}, r_{d}$, respectively, with $r_{d} \ll r_{a}$. The distance between the centers of the particles is $R$, with $R>r_{a}$. The dielectric constant of the surrounding environment is $\varepsilon_{env}$ and is taken to be real. A general description of the quantum dot system including exciton and biexciton transitions is by a four-level system
\cite{Hui08a,Machnikowski08a}. The system is
comprised of a ground state $|0\rangle$, the single-exciton
states with specific circular polarizations $(|\sigma_{+}\rangle,
|\sigma_{-}\rangle)$ and the biexciton state $|2\rangle$. The
single-exciton state $|\sigma_{+}\rangle$ $(|\sigma_{-}\rangle)$
is coupled with a right-handed (left-handed) circularly polarized
field $\sigma_{+}$ $(\sigma_{-})$ to the ground state and with a
left-polarized (right-polarized) field to the biexciton state. An
alternative picture, that is quite useful when linearly polarized fields are applied, is the
description of the single-exciton states with the so-called
linearly polarized states $|X\rangle$ ($x$-polarized) and
$|Y\rangle$ ($y$-polarized), that are written as superpositions of
the states $|\sigma_{+}\rangle, |\sigma_{-}\rangle$, as
$|X\rangle = (|\sigma_{+}\rangle + |\sigma_{-}\rangle)/\sqrt{2}$
and $|Y\rangle = (|\sigma_{+}\rangle -
|\sigma_{-}\rangle)/\sqrt{2}$.

Here, we apply a linearly polarized field along the $x$-axis that induces excitation
in the transition $|0\rangle \rightarrow |X\rangle \rightarrow
|2\rangle$, while the state $|Y\rangle$ remains uncoupled. For concreteness we will rename state $|X\rangle$ to state $|1\rangle$, so the states of coupled system are $|0\rangle$, $|1\rangle$, and $|2\rangle$.  This
way the biexciton system, under the interaction with the
electromagnetic field, has the structure of a three-level
ladder-type (or cascade) system, as it is shown in Fig.\ \ref{fig1}. The linearly polarized oscillating field with $\vec{{\cal E}}(t) = \hat{x} {\cal E}_{0} f(t)\cos(\omega t)$ excites the ground-exciton and the exciton-biexciton transitions in the quantum dot. Note that $\hat{x}$ denotes the unit vector of polarization, ${\cal E}_{0}$ the amplitude and $f(t)$ the envelope (dimensionless) of the electromagnetic pulse, while $\omega$ the field frequency.
The dielectric constants of the quantum dot and the classical metal nanoparticle are $\varepsilon_{S},\varepsilon_{m}(\omega)$, respectively. The applied field also excites plasmons, on the surface of the metal nanoparticle. These plasmonic excitations provide a strong continuous spectral response. The surface plasmons, excited on the nanoparticle due to the applied field, give a powerful continuous spectral response and interact with the quantum dot excitons \cite{Wang06a,Cheng07a,Yan08a,Malyshev15a}, coupling the two particles and leading to F\"{o}rster transfer of energy \cite{Sadeghi13a}. The following Hamiltonian describes the hybrid nanostructure under consideration,
\begin{eqnarray}
H &=& {E}|1\rangle \langle 1| + (2{ E}+{E_{B}})|2\rangle \langle 2|
\nonumber \\ &&- \mu {\cal E_{SQD}}(t)\big{(}|0\rangle
\langle 1| + |1\rangle \langle 2| + H.c. \big{)} \, ,
\label{ham1}
\end{eqnarray}
where we have used the dipole approximation. Here, ${E}$ is the energy of the exciton state and $E_{B}$ is the biexciton energy shift, while we have set the
ground state energy to zero. Also, $\mu$ represents the dipole moment for both the ground-exciton and exciton-biexciton transitions, and ${\cal E_{SQD}}$ represents the electric field inside the quantum dot. For a symmetric quantum dot that we assume here, there exists no direct coupling between the ground and biexciton states,  because of the selection rules.

Under the dipole approximation the total field inside the quantum dot has two parts, the external field and the induced field by the polarization of the classical spherical metallic nanoparticle. The proper calculation of ${\cal E_{QSD}}$ is performed by separating out the positive and negative frequency parts, since they present different time response, thus we write using the quasistatic approximation \cite{Wang06a,Yan08a,Artuso10a,Malyshev15a}:
\begin{equation}
{\cal E_{SQD}}(t) = \frac{\hbar}{\mu}\bigg[\frac{\Omega(t)}{2}e^{-i\omega t} + G \left[\rho_{10}(t)+\rho_{21}(t)\right] + H.c. \bigg] \, , \label{esqd}
\end{equation}
where $\rho_{ij}(t)$ are the elements of density matrix. We also defined the time-dependent Rabi frequency $\Omega(t)$ as \cite{Wang06a,Yan08a,Artuso10a}
\begin{equation}
\Omega(t) = \Omega_{0}f(t) \quad , \quad \Omega_{0} = \frac{\mu {\cal E}_{0}}{\hbar \varepsilon_{effS}}\left(1 + \frac{s_{a}\gamma_{1}r^{3}_{a}}{R^{3}}\right) \, , \label{rabif}
\end{equation}
and parameter $G$ as \cite{Yan08a}
\begin{equation}
G =  \sum^{N}_{n=1}\frac{1}{4\pi\varepsilon_{env}}\frac{(n+1)^2\gamma_{n}r^{2n+1}_{a}\mu^{2}}{\hbar \varepsilon^{2}_{effS} R^{2n+4}} \, . \label{gpar}
\end{equation}
Here, $\varepsilon_{effS} = \frac{2\varepsilon_{env}+\varepsilon_{S}}{3\varepsilon_{env}}$, $\gamma_{n} = \frac{\varepsilon_{m}(\omega) - \varepsilon_{env}}{\varepsilon_{m}(\omega)+(n+1)\varepsilon_{env}/n}$ with $n$ positive integer, while $s_{a} = 2$ because we apply the external field along the interparticle $x$-axis of the nanostructure.

The expression (\ref{rabif}) for the Rabi frequency contains two parts, one which is attributed directly to the external field, and another attributed to the electric field of the metallic nanoparticle, which is actually induced by the applied field. Additionally, parameter $G$ emerges because of the excitons-plasmons interaction \cite{Wang06a,Cheng07a,Yan08a,Miri18a}. Specifically, the external electric field induces a dipole on the quantum dot, which subsequently induces another dipole on the metallic nanoparticle, affecting the quantum dot through the self-interaction parameter $G$ \cite{Wang06a,Artuso10a}. The formula of Eq.\ (\ref{gpar}) provides more realistic values of $G$ since it accounts for multipole effects \cite{Yan08a}. We note that the Hamiltonian given in Eq.\ (\ref{ham1}) and the electric field given in Eq.\ (\ref{esqd}) are similar to those describing a nanostructure consisting of a symmetric quantum dimer emitter and a metallic nanoparticle \cite{Malyshev15a}. Therefore, the present work can find application to that system as well.

The differential equations describing the time evolution of the system, are given by:
\begin{widetext}
\begin{subequations}
\begin{eqnarray} \label{dm}
\dot{\rho}_{00}(t) &=& \Gamma_{11}\rho_{11}(t)+i \frac{\mu {\cal E_{SQD}}(t)}{\hbar}\left[\rho_{10}(t)-\rho_{01}(t)\right] \, ,  \\
\dot{\rho}_{22}(t) &=& -\Gamma_{22}\rho_{22}(t)+i \frac{\mu {\cal E_{SQD}}(t)}{\hbar}\left[\rho_{12}(t)-\rho_{21}(t)\right] \, , \\
\dot{\rho}_{01}(t) &=&\left(i\frac{{E}}{\hbar}-\gamma_{01}\right)\rho_{01}(t)+ i\frac{\mu {\cal E_{SQD}}(t)}{\hbar}\left[\rho_{11}(t)-\rho_{00}(t)\right]- i\frac{\mu {\cal E_{SQD}}(t)}{\hbar}\rho_{02}(t) \, , \\
\dot{\rho}_{02}(t) &=&\left(i\frac{2E+E_{B}}{\hbar}-\gamma_{02}\right)\rho_{02}(t)+ i\frac{\mu {\cal E_{SQD}}(t)}{\hbar}\rho_{12}(t)-i\frac{\mu {\cal E_{SQD}}(t)}{\hbar}\rho_{01}(t) \, , \\
\dot{\rho}_{12}(t) &=&\left(i\frac{E+E_{B}}{\hbar}-\gamma_{12}\right)\rho_{12}(t)+ i\frac{\mu {\cal E_{SQD}}(t)}{\hbar}\left[\rho_{22}(t)-\rho_{11}(t)\right]+ i\frac{\mu {\cal E_{SQD}}(t)}{\hbar}\rho_{02}(t) \, ,
\end{eqnarray}
\end{subequations}
\end{widetext}
where $\sum_{i=1}^{3}\rho_{ii}(t)=1$ and $\rho_{nm}(t)=\rho^{*}_{mn}(t)$. In the above equations $\Gamma_{11}$,
$\Gamma_{22}$ denote the decay rates for the exciton and
biexciton energy levels, respectively, and $\gamma_{nm}$, with $n \neq m$ the dephasing rates of the system. We proceed with a change of variables $\rho_{nn}(t)=\sigma_{nn}(t)$, $\rho_{01}(t) = \sigma_{01}(t)e^{i \omega t}$, $\rho_{02}(t) = \sigma_{02}(t)e^{2 i \omega t}$, and $\rho_{12}(t) = \sigma_{12}(t)e^{i \omega t}$ and make the rotating wave approximation, in order to obtain the time evolution for the slowly varying envelopes of the density matrix elements:
\begin{widetext}
\begin{subequations}
\label{density matrix}
\begin{eqnarray}
\dot{\sigma}_{00}(t) &=& \Gamma_{11}\sigma_{11}(t)+i\frac{\Omega^{*}(t)}{2}\sigma_{10}(t) - i\frac{\Omega(t)}{2}\sigma_{01}(t) \nonumber \\  &+& i G^{*}\left[\sigma_{01}(t)+\sigma_{12}(t)\right]\sigma_{10}(t) - iG\left[\sigma_{10}(t)+\sigma_{21}(t)\right]\sigma_{01}(t) \, , \label{eq1} \\
\dot{\sigma}_{22}(t) &=& -\Gamma_{22}\sigma_{22}(t)+i\frac{\Omega(t)}{2}\sigma_{12}(t) - i\frac{\Omega^{*}(t)}{2}\sigma_{21}(t) \nonumber \\  &+& i G\left[\sigma_{10}(t)+\sigma_{21}(t)\right]\sigma_{12}(t) - iG^{*}\left[\sigma_{01}(t)+\sigma_{12}(t)\right]\sigma_{21}(t) \, , \\
\dot{\sigma}_{01}(t) &=&\left(i\frac{{E}}{\hbar}-i\omega-\gamma_{01}\right)\sigma_{01}(t)+i\frac{\Omega^{*}(t)}{2}\left[\sigma_{11}(t)-\sigma_{00}(t)\right]- i \frac{\Omega(t)}{2}\sigma_{02}(t) \nonumber \\  &+&
iG^{*}\left[\sigma_{01}(t)+\sigma_{12}(t)\right]\left[\sigma_{11}(t)-\sigma_{00}(t)\right]-i G\left[\sigma_{10}(t)+\sigma_{21}(t)\right]\sigma_{02}(t) \, , \\
\dot{\sigma}_{02}(t) &=&\left(i\frac{2{E}+{E_{B}}}{\hbar}-2i\omega-\gamma_{02}\right)\sigma_{02}(t)+i\frac{\Omega^{*}(t)}{2}\left[\sigma_{12}(t)-\sigma_{01}(t)\right]\nonumber \\  &+& iG^{*}\left[\sigma^{2}_{12}(t)-\sigma^{2}_{01}(t)\right] \, , \\
\dot{\sigma}_{12}(t) &=&\left(i\frac{{E+E_{B}}}{\hbar}-i\omega-\gamma_{12}\right)\sigma_{12}(t)+i\frac{\Omega^{*}(t)}{2}\left[\sigma_{22}(t)-\sigma_{11}(t)\right]+ i \frac{\Omega(t)}{2}\sigma_{02}(t) \nonumber \\  &+&
iG^{*}\left[\sigma_{01}(t)+\sigma_{12}(t)\right]\left[\sigma_{22}(t)-\sigma_{11}(t)\right]+i G\left[\sigma_{10}(t)+\sigma_{21}(t)\right]\sigma_{02}(t) \label{eq2} \, .
\end{eqnarray}
\end{subequations}
\end{widetext}

As we already mentioned above, our study is based on the use of the nonlinear density matrix equations presented above for describing the interaction of light and matter in the quantum dot in the presence of the metallic nanosphere. The nonlinear density matrix equations under the quasistatic approximation, including, mainly, exciton effects, and in certain cases \cite{Nugroho19a,Malyshev15a} biexciton effects, have been widely used in literature for the study of optical effects in quantum dots near metallic nanostructures, see, e.g., Refs.\ \cite{Cheng07a,Sadeghi09a,Sadeghi10a,Malyshev13a,Carreno18a,Anton12a,Anton13a,Paspalakis13a,Lee15a,McMillan16a,Qi19a,Smbonias21a,Wang06a,Yan08a,Miri18a,Artuso10a,Malyshev11a,Hatef12a,Kosionis13a,Zhao14a,Singh16a,Kosionis2019,You2019,Singh20a,Nugroho19a,Malyshev15a,Sadeghi13a,Hatef13a}. This method is used in relatively small metal nanoparticles and for distances between the quantum dot and the metal nanoparticle up to approximately half the radius of the metal nanoparticle. However, there is another widely used methodology for studying light-matter interaction in quantum systems near plasmonic nanostrucrures which is based on the electromagnetic Green's tensor, see, e.g., Refs.\ \cite{QED1,Ge2013,Chen13a,Delga14a,Carreno2016a,Carreno19a}, which uses linear density matrix equations, as the Green's tensor does not lead to a direct nonlinearity in the density matrix equations.
One may induce indirectly a nonlinearity in the linear density matrix equations of the electromagnetic Green's tensor approach when a self-consistent scheme is considered and the local field correction to the electric field interacting with the quantum system is introduced, as it was shown by Hayati {\it et al.} \cite{Hayati16a}. In that work a comparison between the two methods was presented and a very good agreement was found \cite{Hayati16a}. We note that the local field correction to the electric field interacting with the quantum dot has been also used in other methodologies involving density matrix equations for the study of light-matter interaction of quantum systems near plasmonic nanostructures, like the finite-difference time-domain method \cite{Lopata09a,Sukharev12a} and the semiclassical (Maxwell-Bloch) model \cite{Gray13a}.

\section{Analytical and numerical results}

Assuming that the quantum dot is prepared in its ground state, the initial values of the density matrix elements are $\sigma_{00}(t=0) = 1$ and $\sigma_{nm}(t=0) = 0$ for the rest. We consider first Eqs. (\ref{density matrix}) for pulses with short durations, so the population decay and dephasing rates can be taken equal to zero, putting thus emphasis on the coherent evolution. The electromagnetic field in the rest of the article is taken at two-photon resonance with the ground-biexciton transition, thus $\hbar\omega = E + E_{B}/2$. In the following we consider that a hyperbolic secant pulse is applied in the system, $f(t) = \mbox{sech}[(t-t_{0})/t_{p}]$, where $t_{p}$ determines the width of the pulse and $t_{0}$ its center, selected so the pulse practically vanishes for $t=0$ and $t=2t_{0}$.

\subsection{$E_{B} = 0$}
Here we study the special case that the biexciton energy shift ${E_{B}}$ is zero. In this case, there is simultaneous single-photon and two-photon resonance. Assuming that $G=0$, we can select the applied field amplitude as the following function of interparticle distance $R$, see appendix \ref{appenA1},
\begin{equation}
{\cal E}_{0} = \frac{\sqrt{2}\hbar\varepsilon_{effS}}{\mu t_{p}\left|1 + \frac{s_{a}\gamma_{1}r^{3}_a}{R^{3}}\right|} \, , \label{pipulse}
\end{equation}
and obtain the analytical solution
\begin{subequations}
\begin{eqnarray}
\sigma_{00}(t) &=& \frac{1}{4} \left[1-\mbox{tanh}\left(\frac{t-t_{0}}{t_{p}}\right)\right]^2 \, , \label{anal0} \\
\sigma_{22}(t) &=& \frac{1}{4} \left[1+\mbox{tanh}\left(\frac{t-t_{0}}{t_{p}}\right)\right]^2 \, . \label{anal1}
\end{eqnarray}
\end{subequations}
Then, complete transfer to the biexciton state is succeeded at the pulse end.

\begin{figure}[ht]
\centering
$\begin{array}{cc}
\includegraphics[width=3in]{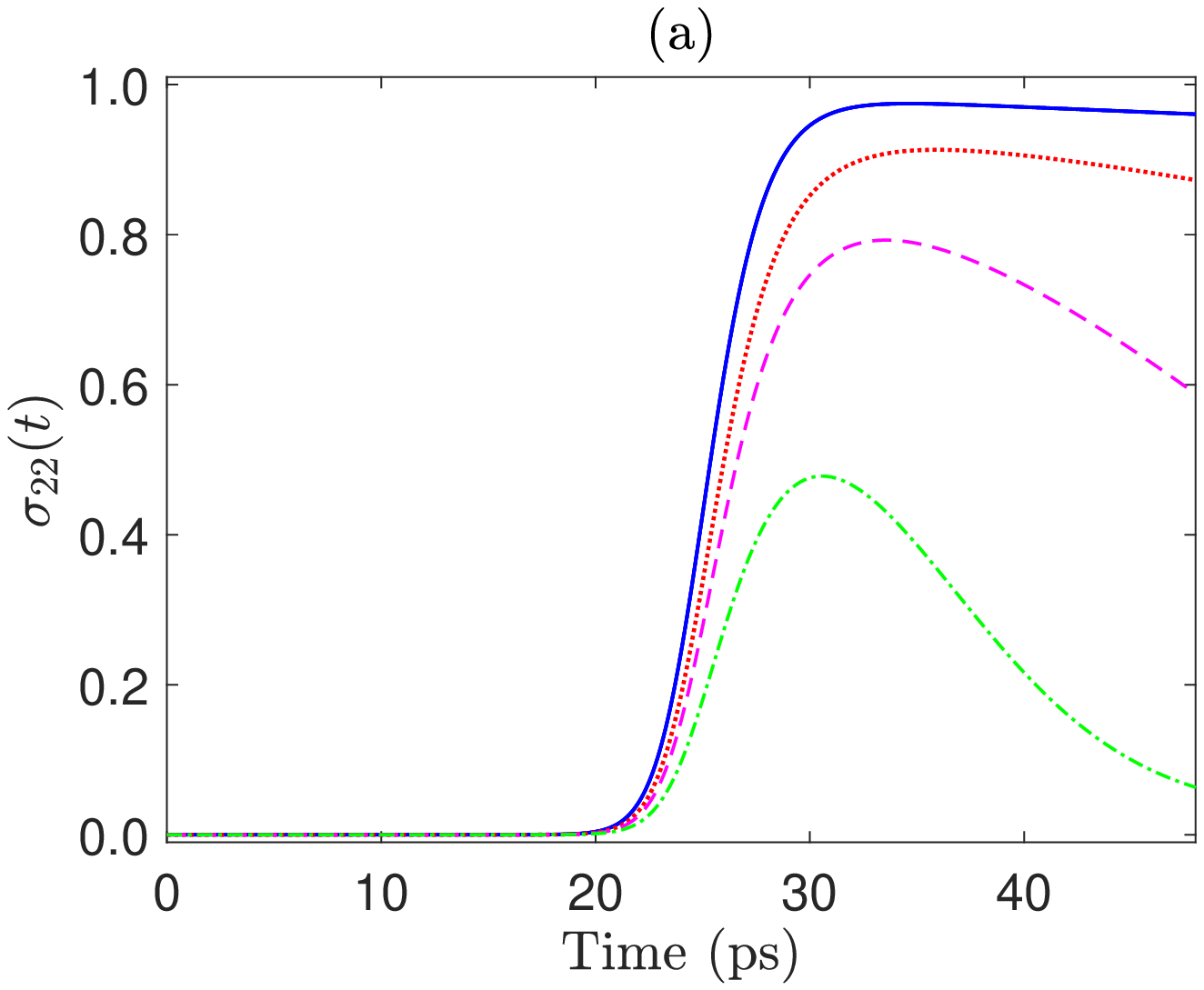} \\
\includegraphics[width=3in]{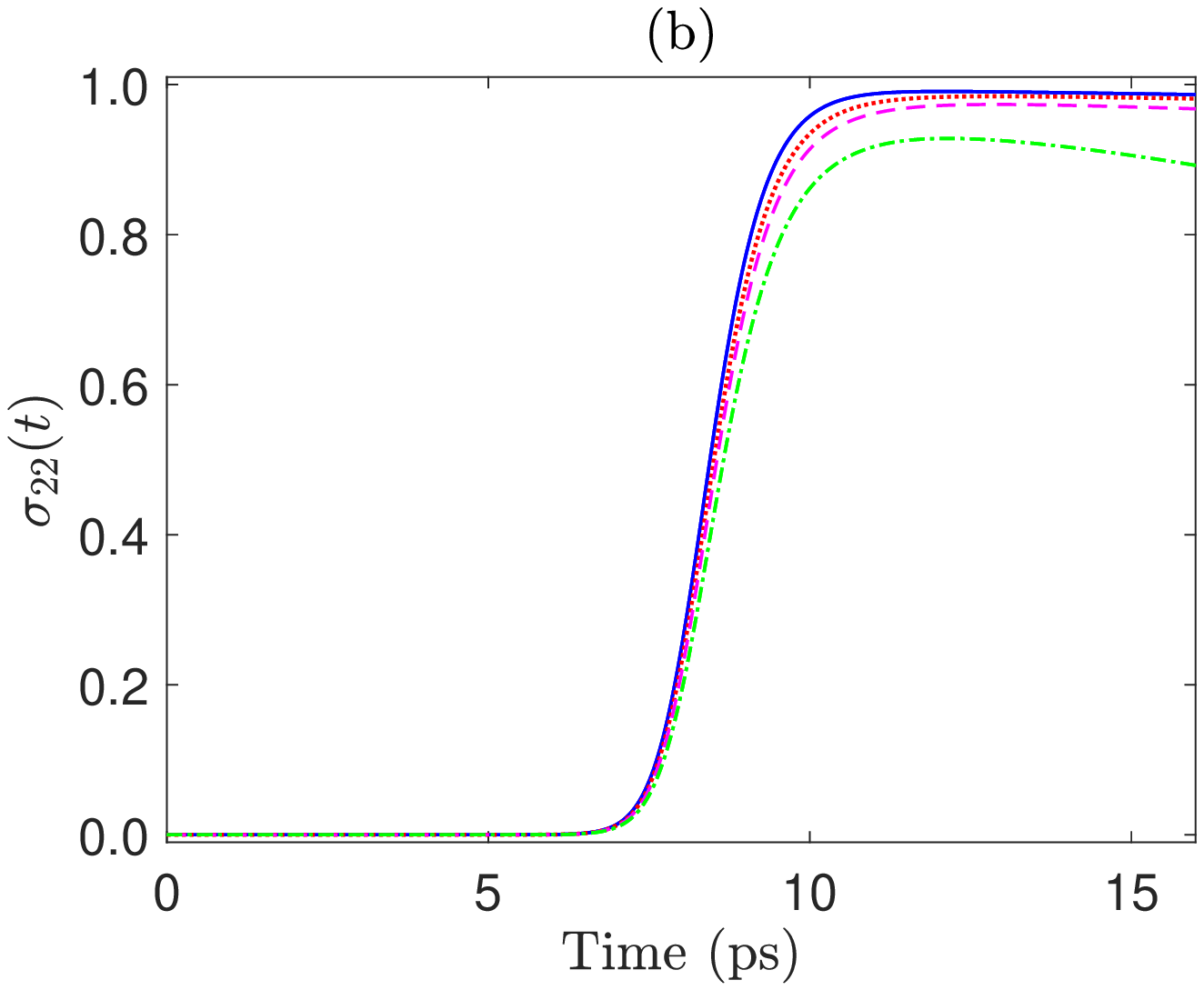} \\
\includegraphics[width=3in]{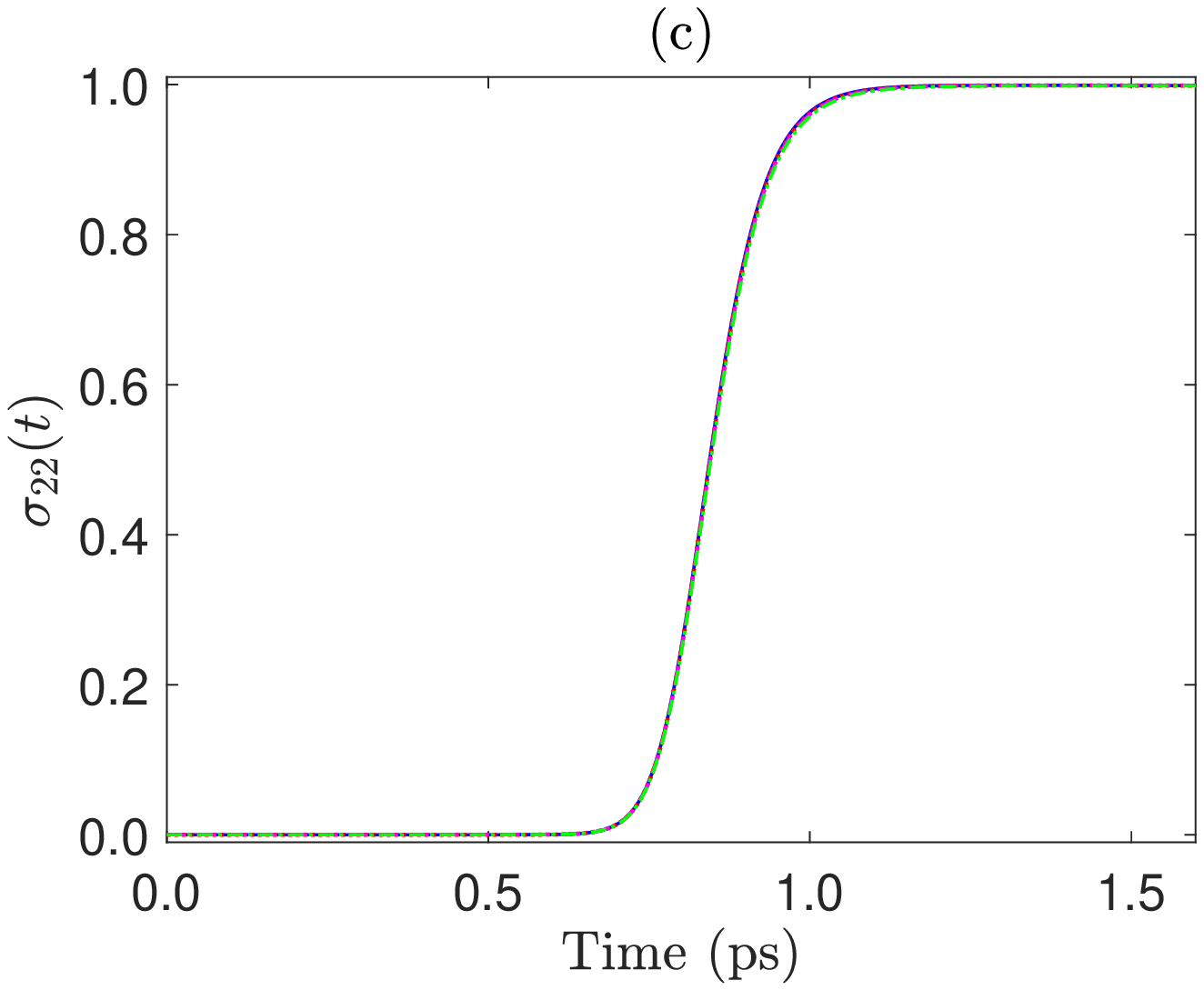}
\end{array}$
\caption{Biexciton population $\sigma_{22}(t)$ obtained by numerically solving system (\ref{density matrix}) for $R =$ 80 nm (blue solid curve), $R =$ 15 nm (red dotted curve), $R =$ 14 nm (magenta dashed curve), and $R =$ 13 nm (green dot-dashed curve). In (a) $t_{p} = 3$ ps, (b) $t_{p} =1$ ps, and (c) $t_{p} =0.1$ ps. In all cases $t_{0} = 8t_{p}$. The applied field amplitude ${\cal E}_{0}$ for each distance is obtained from Eq.\ (\ref{pipulse}).} \label{fig2}
\end{figure}

The metallic nanoparticle can strongly affect the time evolution of populations for small $R$ and considerably reduce the population transferred
to the biexciton state. In Fig.\ \ref{fig2} we display $\sigma_{22}(t)$ as obtained from the numerical simulation of the full system Eqs. (\ref{density matrix}). For the simulation, we use the parameter values $\Gamma^{-1}_{11} = \Gamma^{-1}_{22} =0.8$ ns, $\gamma^{-1}_{01}=\gamma^{-1}_{02}=\gamma^{-1}_{12}$ = 0.3 ns, $\varepsilon_{env} = \varepsilon_{0}$, $r_{a} = 7.5$ nm, $\mu = 0.65$ $e$ nm, $\hbar\omega_{0} = 2.5$ eV, and $\varepsilon_{S} = 6\varepsilon_{0}$, where $\varepsilon_{0}$ is the vacuum dielectric constant. They are typical values for colloidal quantum dots, for example CdSe-based, and are also utilized in several other studies \cite{Wang06a,Yan08a,Sadeghi09a,Sadeghi10a,Artuso10a,Malyshev11a,Paspalakis13a,Koch10a}. Regarding the dielectric constant of the nanoparticle, $\varepsilon_{m}(\omega)$, we use experimental data for gold \cite{Johnson72a}. We also use a nonzero self-interaction parameter $G$. Observe from Eq. (\ref{gpar}) that $G$ increases when the distance $R$ decreases, thus affecting more the evolution of populations and rendering solution (\ref{anal1}) incorrect. This is clearly illustrated in Fig.\ \ref{fig2}(a)

Similar to the case where only the population dynamics of the exciton state is considered \cite{Paspalakis13a}, the effect of parameter $G$ on the biexciton population, too, is considerably determined by the duration of the pulse, as illustrated in Figs.\ \ref{fig2}(b) and \ref{fig2}(c). For shorter pulses, the effect of $G$ significantly weakens [see Fig.\ \ref{fig2}(b)], and when the duration becomes very small no influence can be observed [see Fig.\ \ref{fig2}(c)]. Thus, an electric field of the form of Eq.\ (\ref{pipulse}) may lead to significant population transfer to the biexciton state, when pulses with shorter durations are used. Additionally, for shorter durations the effect of decay and dephasing is also reduced, leading to higher biexciton population at the final time.

\begin{figure}[ht]
\centering
$\begin{array}{cc}
\includegraphics[width=3in]{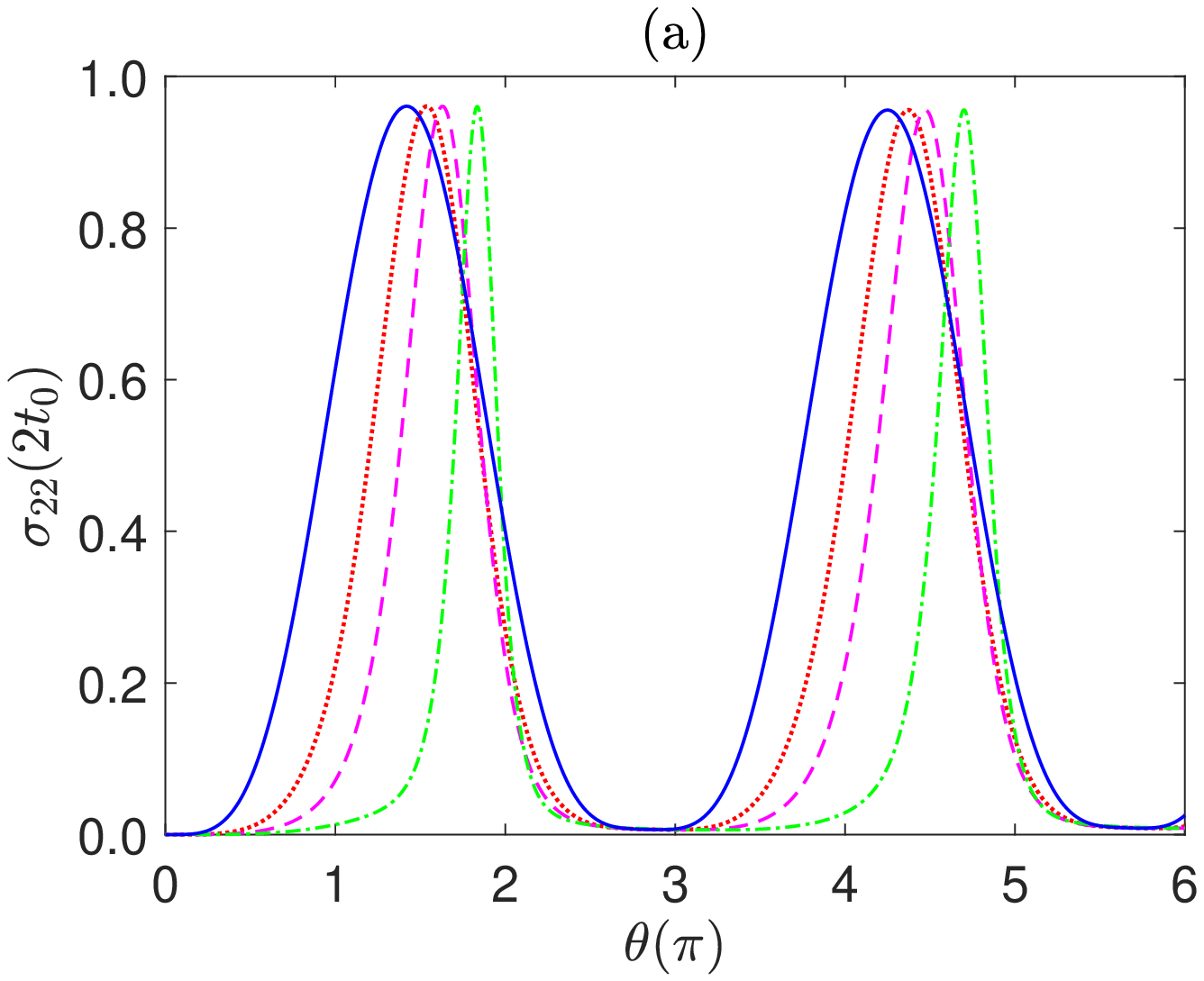} \\
\includegraphics[width=3in]{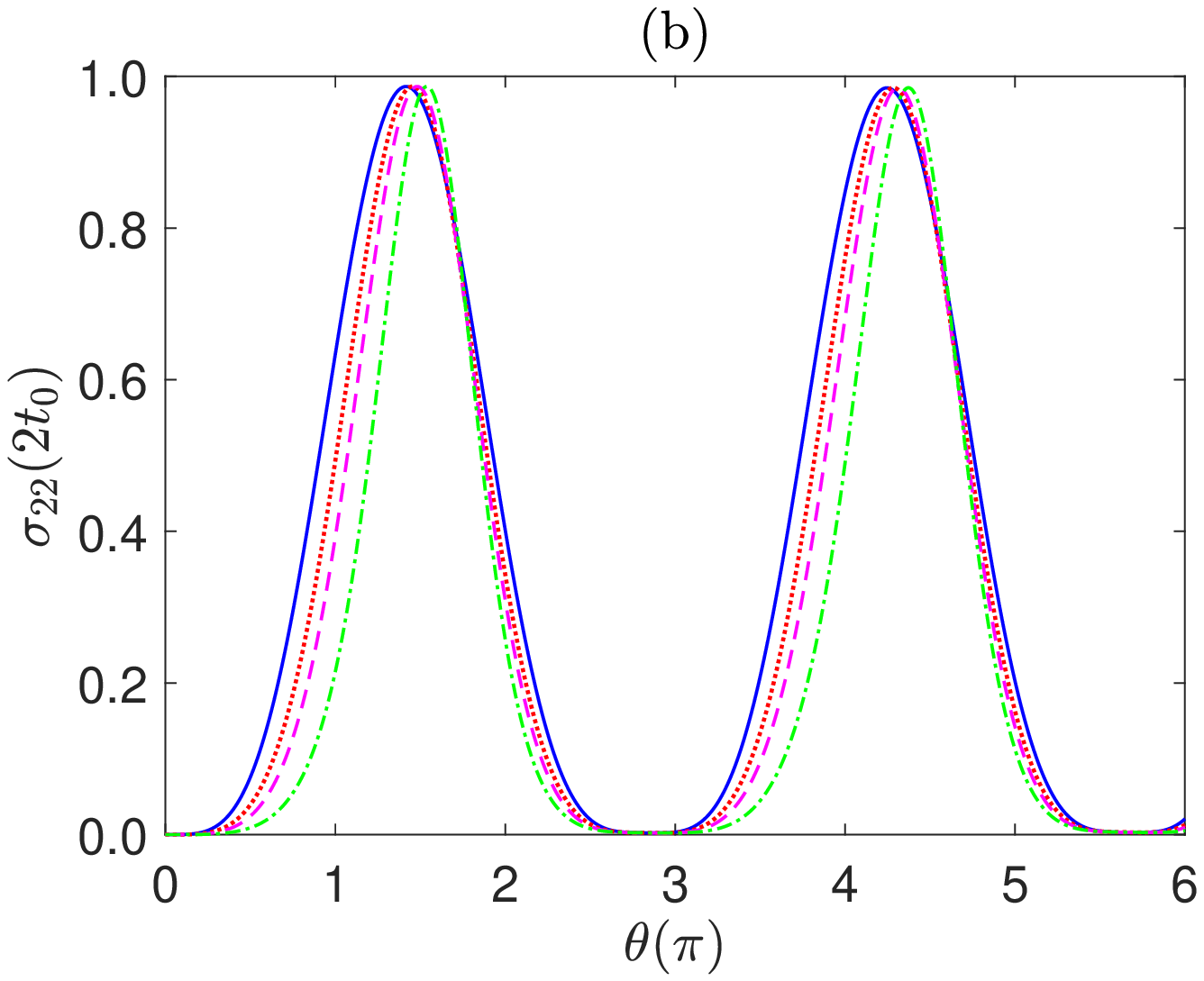} \\
\includegraphics[width=3in]{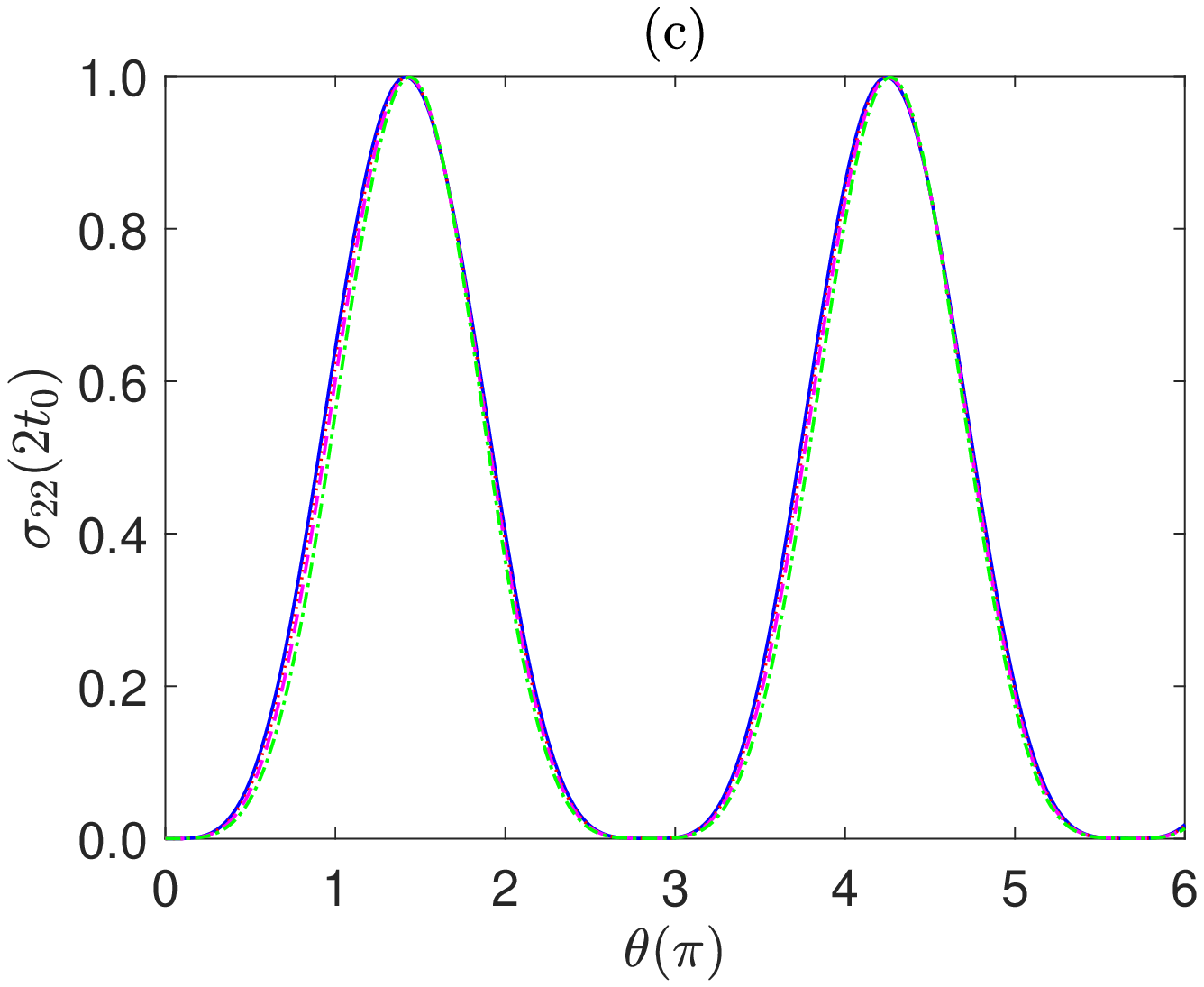}
\end{array}$
\caption{Biexciton population at the final time $t=2t_0$, $\sigma_{22}(2t_{0})$, versus the pulse area $\theta$ (taken in multiples of $\pi$). In (a) $t_{p} = 3$ ps, (b) $t_{p} =1$ ps, and (c) $t_{p} =0.1$ ps. In all cases $t_{0} = 8t_{p}$. The applied field amplitude ${\cal E}_{0}$ for each distance is obtained from Eq.\ (\ref{pipulse}). Blue solid curve: $R =$ 80 nm, red dotted curve: $R =$ 15 nm, magenta dashed curve: $R =$ 14 nm, and green dot-dashed curve: $R =$ 13 nm. } \label{fig3}
\end{figure}

We investigate further how the population transfer depends on the system parameters and display in Fig.\ \ref{fig3} the final value of the population in the biexciton state $\sigma_{22}(t)$, i.e., the value of $\sigma_{22}(t)$ at $t=2t_{0}$, versus the pulse area $\theta$ [absolute value of the time integral of Eq.\ (\ref{rabif}) for the complete pulse duration, which gives $\theta = \pi |\Omega_{0}|t_{p}$]. Note that we obtain these results by numerically integrating Eqs.\ (\ref{eq1})-(\ref{eq2}) with nonzero $G$. Observe that for longer pulse durations, the areas corresponding to maximum population transfer can become larger than $\sqrt{2}\pi$, which is the value that gives the maximum population inversion, ideally complete transfer for zero decay and dephasing effects, when the metallic nanoparticle is not present. Additionally, the actual value of pulse area that gives the maximum transfer for each case depends strongly on $R$, with larger pulse areas required for shorter interparticle distances, where the influence of the nanoparticle is stronger. This point sheds light to the results of Fig.\ \ref{fig2}, since we can observe that for longer pulses the chosen $\sqrt{2}\pi$ pulse works well for long interparticle distances. For shorter pulses, the area giving maximum population transfer approaches $\sqrt{2}\pi$, for several quantum dot-metallic nanoparticle distances. For the shortest pulse duration used, in Fig.\ \ref{fig3}(c), the maximum biexciton population is achieved for area $\sqrt{2}\pi$ and its odd multiples, practically for all values of interparticle distances.

\begin{figure}[ht]
\centering
\includegraphics[width=3.5in]{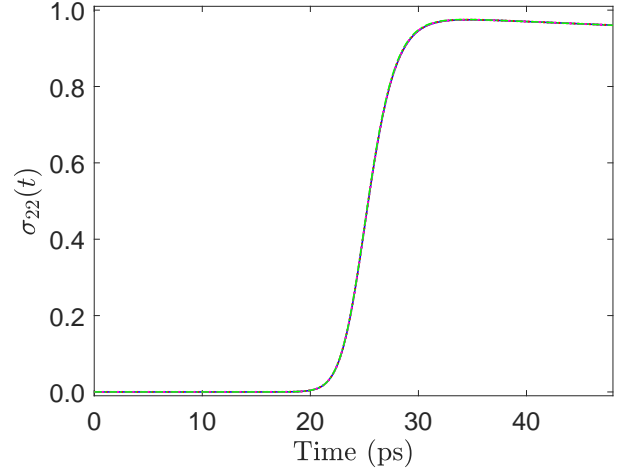}
\caption{Biexciton population $\sigma_{22}(t)$ obtained by numerically solving system (\ref{density matrix}) for $R =$ 80 nm (blue solid curve), $R =$ 15 nm (red dotted curve), $R =$ 14 nm (magenta dashed curve), and $R =$ 13 nm (green dot-dashed curve) with $t_{p} = 3$ ps and $t_{0} = 8t_{p}$. The applied field amplitude ${\cal E}_{0}$ for each distance is obtained from Eq.\ (\ref{pi2}).
} \label{fig4}
\end{figure}

Interestingly, for the case of non-zero $G=G_R+iG_I$, analytical solutions (\ref{anal0}) and (\ref{anal1}) may also be preserved if the electric field amplitude is chosen as
\begin{equation}
{\cal E}_{0} = \frac{\sqrt{2}\hbar\varepsilon_{effS}\left[G_{R}^{2}+\left(G_{I}+\frac{1}{t_{p}}\right)^{2}\right]^{1/2}}{\mu\left|1 + \frac{s_{a}\gamma_{1}r^{3}_{a}}{R^{3}}\right|} \, . \label{pi2}
\end{equation}
For the derivation of Eq.\ (\ref{pi2}), see appendix \ref{appenA1}.
When $G=0$, Eq.\ (\ref{pipulse}) is recovered from Eq.\ (\ref{pi2}).
In order to evaluate the analytical results for this case, in Fig.\ \ref{fig4} is plotted the biexciton population $\sigma_{22}(t)$, found by numerically integrating Eqs. (\ref{density matrix}) using the previous parameter values and several distances, for the same duration $t_{p} = 3$ ps as in Fig.\ \ref{fig2}(a) but with the electric field amplitude given by Eq.\ (\ref{pi2}). Observe that now the curves corresponding to various interparticle distances can be hardly distinguished. Compared to the situation displayed in Fig.\ \ref{fig2}(a), where the electric field amplitude (\ref{pipulse}) is used, for both cases the dissipation and dephasing processes prohibit the complete inversion of population, nevertheless high levels of population transfer to the biexciton state are achieved now for all the displayed interparticle distances. One may use Eq.\ (\ref{pi2}) to understand how the population dynamics behaves for short pulse durations, as shown in Fig.\ \ref{fig2}(c). For specified quantum dot, metal nanoparticle and interparticle distance, the parameters $R$, $G_{I}$ and $G_{R}$, are fixed. Then, shorter pulses have higher values of $1/t_{p}$, and in the limit of small enough $t_p$ it is $\sqrt{G_{R}^{2}+\left(G_{I}+1/{t_{p}}\right)^{2}} \approx 1/t_{p}$, thus Eq.\ (\ref{pipulse}) approximates Eq.\ (\ref{pi2}).

\begin{figure}[ht]
\centering
$\begin{array}{cc}
\includegraphics[width=3in]{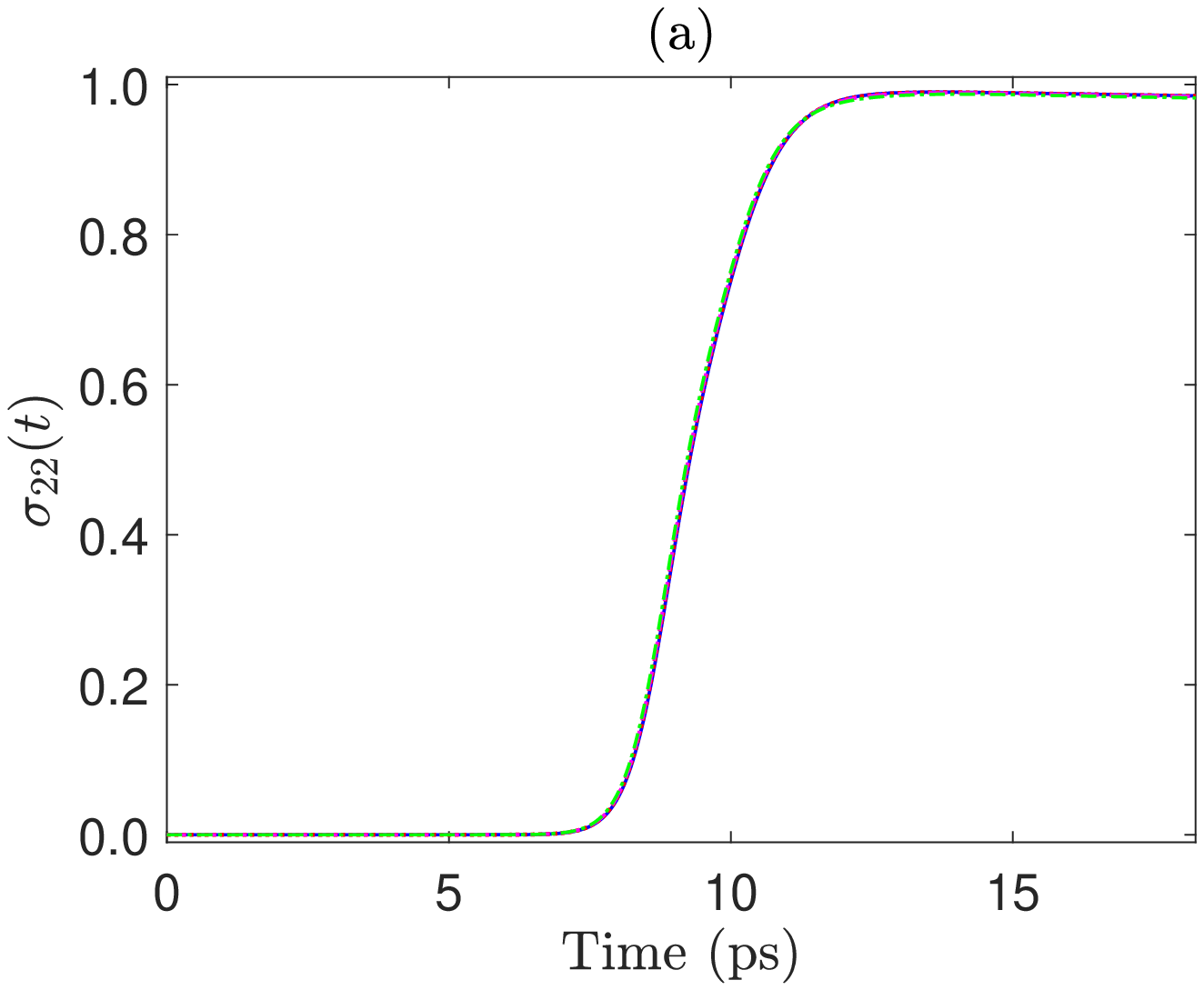} \\
\includegraphics[width=3in]{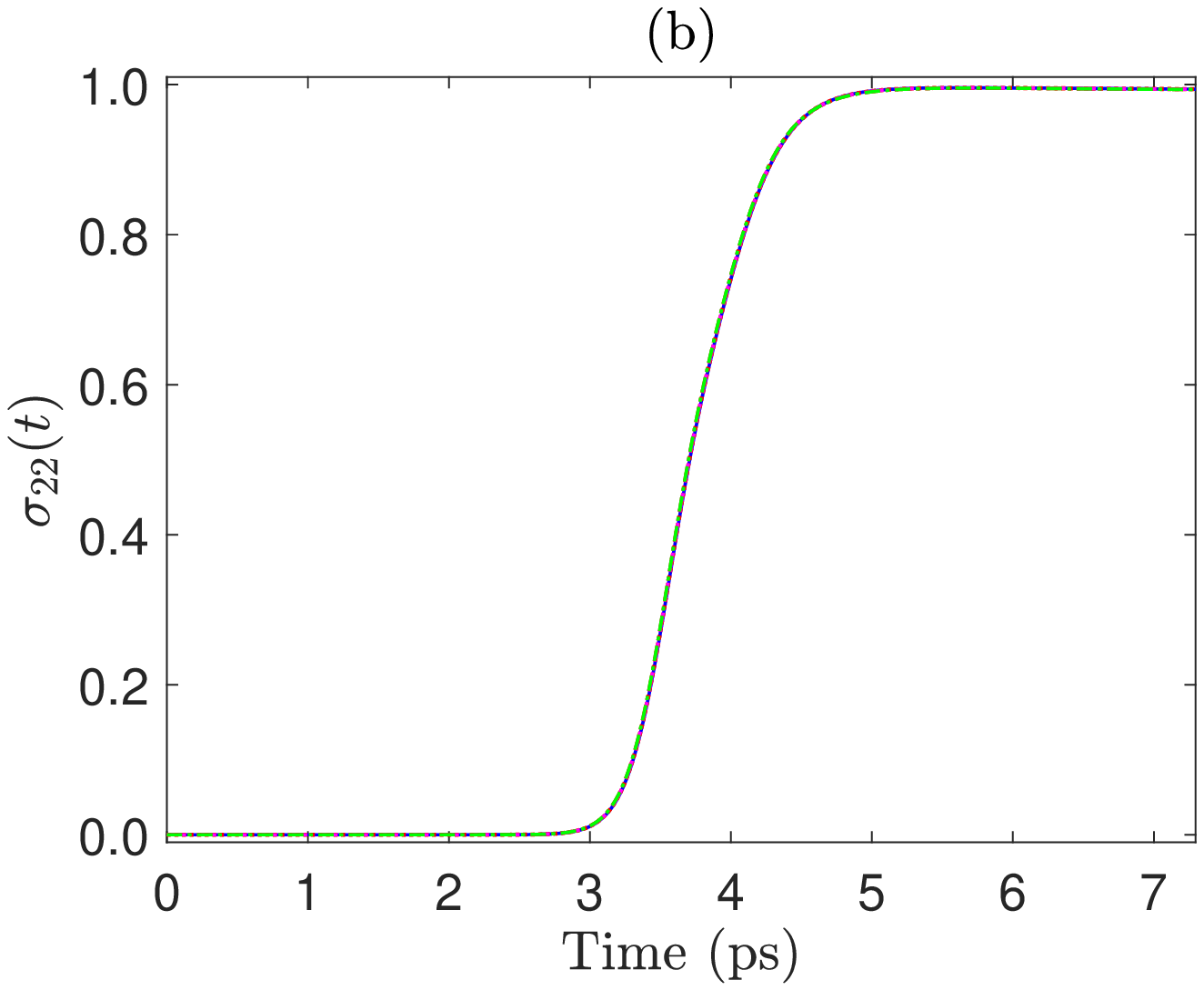} \\
\includegraphics[width=3in]{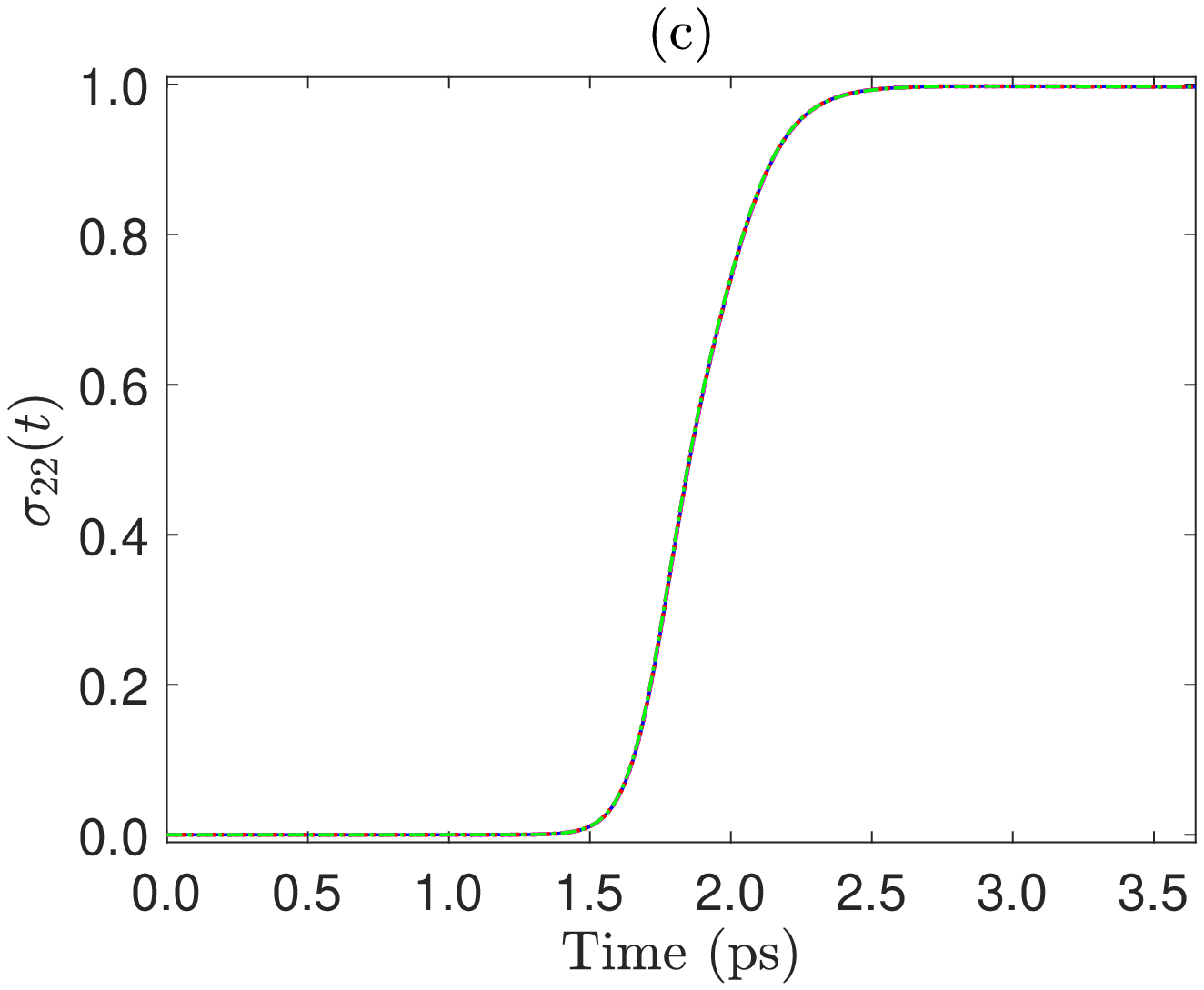}
\end{array}$
\caption{Biexciton population $\sigma_{22}(t)$ obtained by numerically solving system (\ref{density matrix}) for $R =$ 80 nm (blue solid curve), $R =$ 15 nm (red dotted curve), $R =$ 14 nm (magenta dashed curve), and $R =$ 13 nm (green dot-dashed curve). In (a) $E_{B} = -2$ meV, (b) $E_{B} = -5$ meV, and (c) $E_{B} = -10$ meV. The applied field amplitude ${\cal E}_{0}$ for each distance is obtained from Eq.\ (\ref{pi3}) with $n=2$, $t_{p}$ satisfies Eq.\ (\ref{parrosen1a}), and $t_{0} = 8t_{p}$.} \label{fig5}
\end{figure}

\subsection{$E_{B} \neq 0$}
Now we turn to the study  of the more general case that the biexciton energy shift ${E_{B}}$ is nonzero. By taking $G=0$ and then following the method of controlled rotation \cite{Paspalakis10a,Kis04b,Paspalakis04b,Kis04a}, see appendix \ref{appenA2}, we obtain the analytical solutions
\begin{subequations}
\begin{eqnarray}
\sigma_{00}(t) &=& \frac{1}{4} \left|\alpha(t)+1\right|^2 \, , \label{anal2} \\
\sigma_{22}(t) &=& \frac{1}{4} \left|\alpha(t)-1\right|^2 \, , \label{anal3}
\end{eqnarray}
\end{subequations}
with
\begin{equation}
\alpha(t) =  {}_{2}F_{1}\left[l,m;q;z(t)\right] \, , \label{hyper}
\end{equation}
and
\begin{subequations}
\begin{eqnarray}
l &=& -m = \frac{|\Omega_{0}|t_{p}}{\sqrt{2}} \, , \\
q &=& \frac{1}{2}-i\frac{E_{B}t_{p}}{4\hbar} \, , \\
z(t) &=& \frac{1}{2}\left[\mbox{tanh}\left(\frac{t-t_{0}}{t_{p}}\right)+1\right] \, ,
\end{eqnarray}
\end{subequations}
where ${}_{2}F_{1}\left[l,m;q;z(t)\right]$ is the hypergeometric function.

Complete transfer to the biexciton state is succeeded at the pulse end for $\alpha(t\rightarrow\infty)=-1$. This gives
\begin{equation}
|\Omega_{0}| = \sqrt{2}n/t_{p} \, , \, n = 1,2,3,... ,
\end{equation}
leading to the $R$-dependent electric field amplitude that satisfies
\begin{equation}
{\cal E}_{0} = \frac{\sqrt{2}n\hbar\varepsilon_{effS}}{\mu t_{p}\left|1 + \frac{s_{a}\gamma_{1}r^{3}_{a}}{R^{3}}\right|} \, ,  \, n = 1,2,3,... , \label{pi3}
\end{equation}
and the pulse duration $t_{p}$ is given by the solution of the equation
\begin{equation}
(-1)^{n}\prod^{n-1}_{j=0}\frac{j+\frac{1}{2}+i\frac{E_{B}t_{p}}{4\hbar}}{j+
\frac{1}{2}-i\frac{E_{B}t_{p}}{4\hbar}} = -1 \, . \label{eks}
\end{equation}
For example, for $n=2$ the solution of Eq.\ (\ref{eks}) gives
\begin{equation}
t_{p} = 2\sqrt{3}\frac{\hbar}{|E_{B}|} \, , \label{parrosen1a}
\end{equation}
while for $n=3$ the solution of Eq.\ (\ref{eks}) gives
\begin{equation}
t_{p}  = 2\sqrt{23}\frac{\hbar}{|E_{B}|} \, . \label{parrosen2a}
\end{equation}
Therefore, the necessary time for complete population transfer to the biexciton state
increases with $n$ and decreases with increasing $|E_{B}|$.

\begin{figure}[t]
\centering
\includegraphics[width=3.5in]{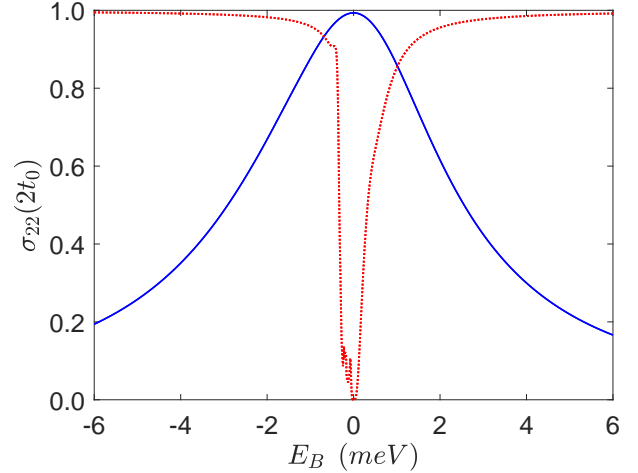}
\caption{Biexciton population at the final time $t=2t_0$, $\sigma_{22}(2t_{0})$, versus $E_{B}$ for $R = 13$ nm. For the blue solid curve the electric field amplitude ${\cal E}_{0}$ satisfies Eq.\ (\ref{pi2}), $t_{p} = 0.5$ ps and $t_{0} = 8t_{p}$. For the red dotted curve the electric field amplitude ${\cal E}_{0}$ satisfies Eq.\ (\ref{pi3}) for $n=2$, $t_{p}$ satisfies Eq.\ (\ref{parrosen1a}), and $t_{0} = 8t_{p}$.} \label{fig6}
\end{figure}

\begin{figure}[ht]
\centering
$\begin{array}{cc}
\includegraphics[width=3in]{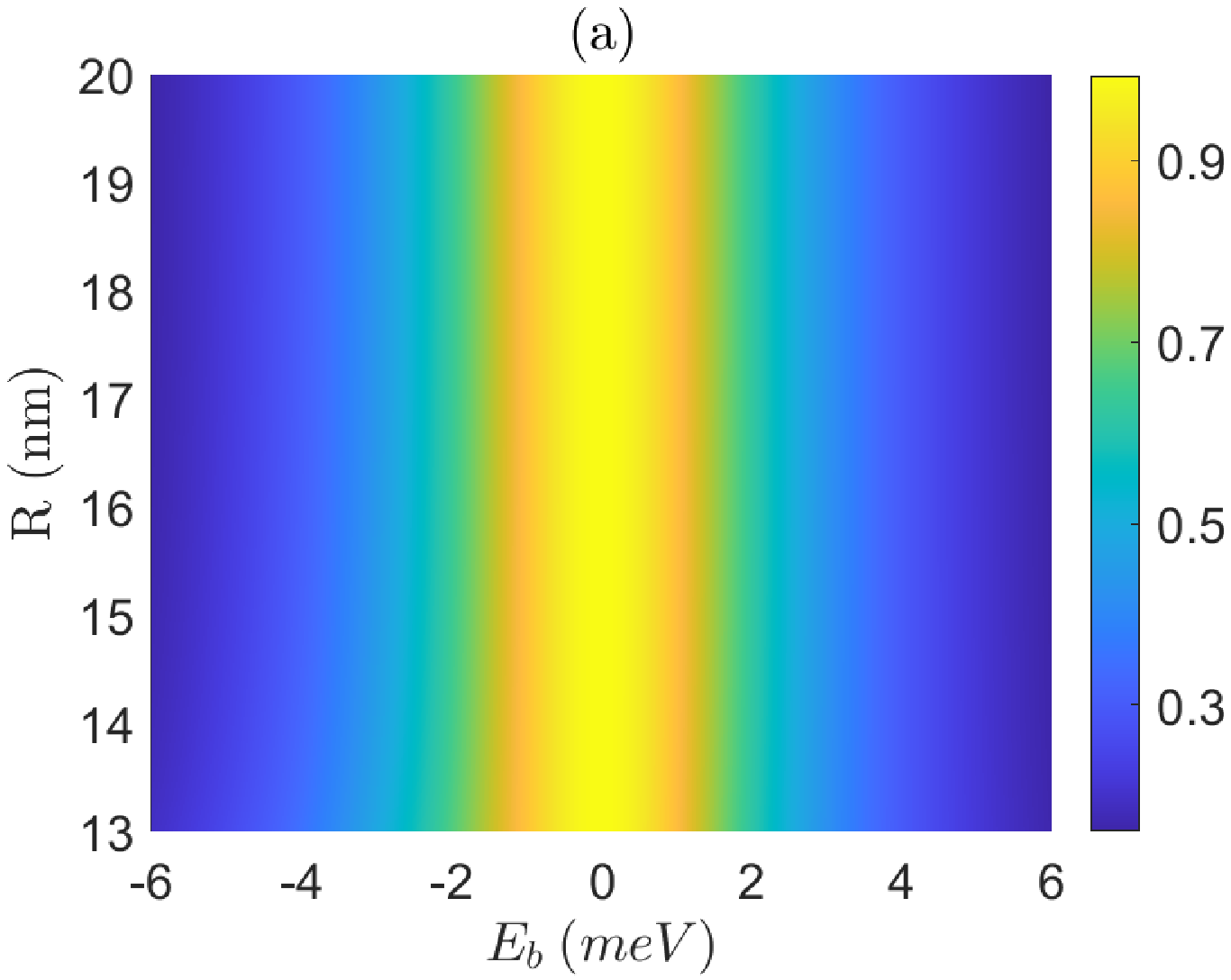} \\
\includegraphics[width=3in]{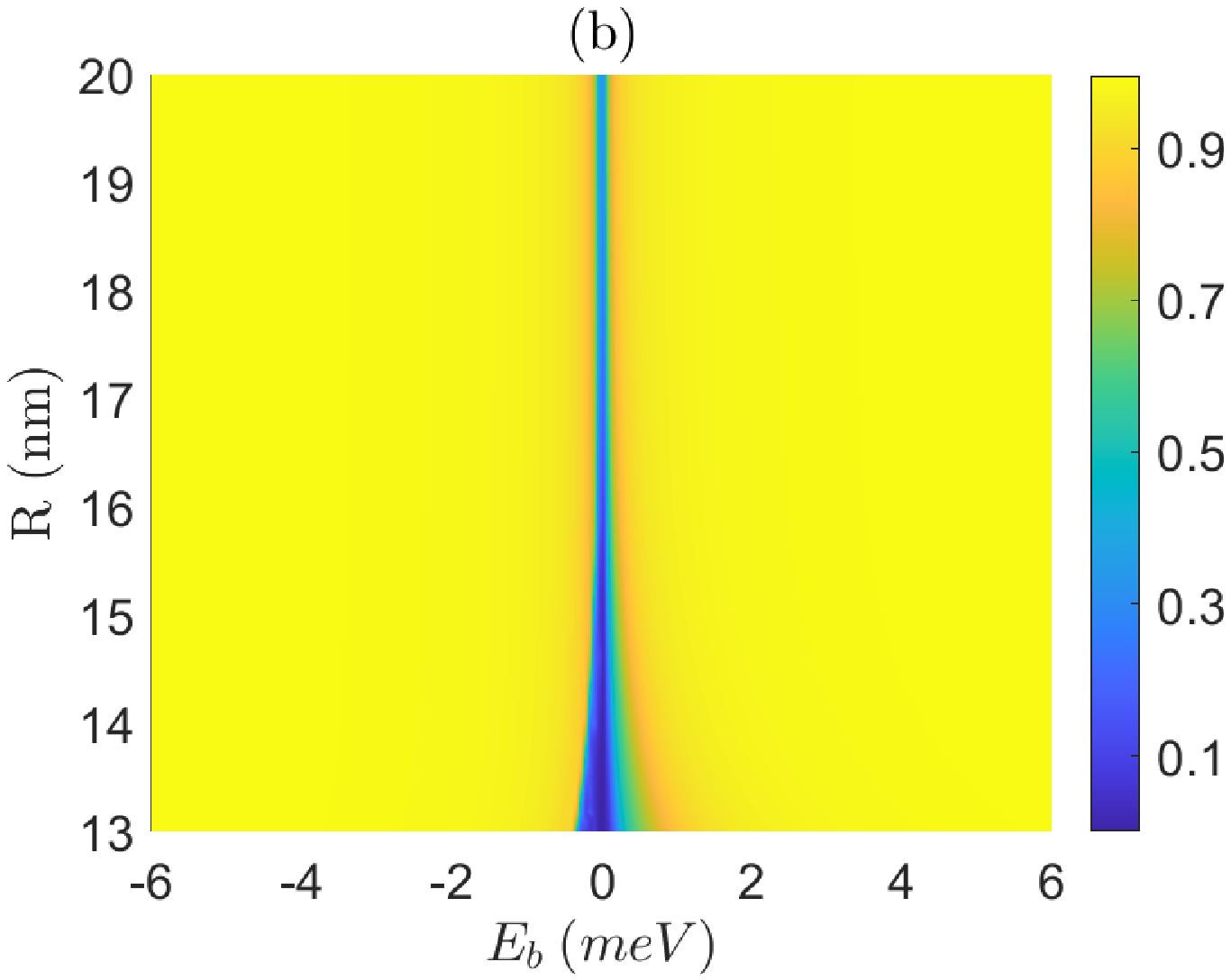}
\end{array}$
\caption{Contour plots of the final values of the biexciton state population $\sigma_{22}(t)$ $[\sigma_{22}(2t_{0})]$ versus the biexciton binding energy $E_B$ and the distance $R$ between the metallic nanoparticle and the semiconductor quantum dot. In (a) the electric field amplitude ${\cal E}_{0}$ satisfies Eq.\ (\ref{pi2}), $t_{p} = 0.5$ ps and $t_{0} = 8t_{p}$. In (b) the electric field amplitude ${\cal E}_{0}$ satisfies Eq.\ (\ref{pi3}) for $n=2$, $t_{p}$ satisfies Eq.\ (\ref{parrosen1a}), and $t_{0} = 8t_{p}$.} \label{fig7}
\end{figure}

The above analytical results are assessed in Fig.\ \ref{fig5}, where we present the biexciton population $\sigma_{22}(t)$, found by numerically integrating Eqs. (\ref{density matrix}) using the previous parameter values, several distances, and different values of the biexciton shift. We emphasize that in the simulations we have included the decay and dephasing rates and the parameter $G$. The electric field amplitude ${\cal E}_{0}$ and $t_{p}$ satisfy Eqs.\ (\ref{pi3}), for $n=2$, and (\ref{parrosen1a}), respectively. For a small value of $|E_{B}|$ shown in Fig.\ \ref{fig5}(a), there is incomplete transfer to the biexciton state due to decay and dephasing processes. However, the efficiency of the transfer is still high, higher than 0.98 for every interparticle distance. The influence of the parameter $G$ is very weak and essentially does not change the efficiency of the transfer since the results seem practically identical for the various interparticle distances, and the corresponding curves are hardly distinguished from each other. For a higher absolute value of the biexciton energy shift, shown in Fig.\ \ref{fig5}(b), very high transfer efficiency to the biexciton state is obtained for all the interparticle distances. For the last case of even higher value of $|E_{B}|$, shown in Fig.\ \ref{fig5}(c), the biexciton state generation is essentially complete. The weak influence of $G$, as well as, of the decay and dephasing rates on the population dynamics with the increase of $|E_{B}|$, occurs due to the decrease of $t_{p}$ and therefore the decrease of the interaction time. We note that CdSe-based quantum dots, as the ones studied here, give in general rather high absolute values of the biexciton shift \cite{Sewall09a}.

We also explore further the dependence of the population transfer to the biexciton level on the biexciton energy shift, by comparing the results of the two conditions, that of Eq.\ (\ref{pi2}), which is obtained for $E_{B} = 0$ and $G \neq 0$, and that of Eqs.\ (\ref{pi3}) and (\ref{parrosen1a}), that are obtained with $E_{B} \neq 0$ and $G=0$. We display in Fig.\ \ref{fig6} the final population of the biexciton state, i.e., the value of $\sigma_{22}(t)$ at $t=2t_{0}$, versus the biexciton energy shift $E_{B}$ for $R=13$ nm. This is obtained from the numerical solution of Eqs.\ (\ref{eq1})-(\ref{eq2}) taking $G$ nonzero and including decay and dephasing terms. We find that for small values of $E_{B}$, typically for $|E_{B}| < 1.5$ meV for the parameters used here, the condition of Eq.\ (\ref{pi2}), derived for $E_{B} = 0$, gives better population transfer than that of Eqs.\ (\ref{pi3}) and (\ref{parrosen1a}), derived for $E_{B} \neq 0$. This happens because Eq.\ (\ref{parrosen1a}) gives large values of $t_{p}$ for small values of $E_{B}$, leading to large interaction times. Therefore, there is very small population transfer in that case to the biexciton state due to the effects of the decay and dephasing rates and the strong influence of $G$. Consequently, the condition of Eq.\ (\ref{pi2}) is preferable for small values of $|E_{B}|$.

For a further assessment of the influence of the interparticle distance $R$ and the biexciton binding energy $E_B$ on the transfer efficiency, we present in Fig.\ \ref{fig7} two-dimensional contour plots of the final biexciton state population $\sigma_{22}(2t_{0})$ versus both variables, when the pulses of Eq.\ (\ref{pi2}) in Fig.\ \ref{fig7}(a) and of Eqs.\ (\ref{pi3}) and (\ref{parrosen1a}) in Fig.\ \ref{fig7}(b) are used. The results are obtained from the numerical solution of Eqs.\ (\ref{eq1})-(\ref{eq2}) taking $G$ nonzero and including decay and dephasing terms. The displayed results confirm those found in Fig.\ \ref{fig6}, that a pulse obtained from Eq.\ (\ref{pi2}) is preferable for small values of biexciton binding energy, while for larger values the pulse obtained from Eqs.\ (\ref{pi3}) and (\ref{parrosen1a}) is favourable. The dependency of the efficiency on the distance $R$ is weak in both cases, and only for short interparticle distances and small values of the biexciton binding energy a small influence on the efficiency of the transfer is found in Fig.\ \ref{fig7}(b).

\section{Conclusion}
In the present work, we analyzed the controlled population transfer to the biexciton state in a hybrid nanostructure composed of a semiconductor quantum dot and a spherical metallic nanoparticle. The quantum dot is modeled as a three-level ladder-type system whose interaction with an externally applied electromagnetic field is described by the modified nonlinear density matrix equations. We consider a pulsed field with hyperbolic secant envelope and derive analytical solutions for the system equations, for both zero and nonzero biexciton energy shift. These solutions lead to efficient transfer to the biexciton state of the quantum dot for a wide range of interparticle distances, which can even have relatively small values. In certain cases, when these distances are small enough, the population transfer is strongly modified because of the interaction of quantum dot excitons with surface plasmons on the nanoparticle. The modification is considerably affected by the duration of the applied pulse. Although here we have analyzed a coupled nanohybrid containing a spherical metal nanoparticle, we note that the presented methodologies can be easily extended in other cases of metallic nanoparticles, such as, for example, for metallic nanorods \cite{Hatef13a}, by simply changing the form of the parameters $\Omega_{0}$ and $G$. The present results may find application in the area of quantum information processing, where efficient preparation of a biexciton state of a semiconductor quantum dot is a useful task.

\section*{DATA AVAILABILITY}

The data that support the findings of this study are available from the corresponding author upon reasonable request.

\section{Acknowledgements}
The work of A. S. is co-financed by Greece and the European Union (European Social Fund-ESF) through the Operational Programme ``Human Resources Development, Education and Lifelong Learning", project ``Strengthening Human Resources Research Potential via Doctorate Research" (MIS-5000432), implemented by the State Scholarships
Foundation (IKY). We acknowledge useful discussions with Prof. Andreas F. Terzis and Prof. Ioannis Thanopulos.

\appendix%
\section{Coherent dynamics for the $E_B = 0$ case} \label{appenA1}
In this part of the Appendix, we present the steps for obtaining the solutions (\ref{anal0}) and (\ref{anal1}).
We take the population decay and dephasing rates as zero in Eqs.\ (\ref{eq1})-(\ref{eq2}). We also define $\sigma(t) = [\sigma_{10}(t)+\sigma_{21}(t))]\sqrt{2}$, $\Delta(t) = \sigma_{00}(t)-\sigma_{22}(t)$, and $\tilde{\Omega}(t) = \Omega(t)/\sqrt{2}$. Using Eqs.\ (\ref{eq1})-(\ref{eq2}) we obtain the following equations for $\sigma(t)$ and $\Delta(t)$ as
\begin{eqnarray}
\dot{\sigma}(t) &=&  i \frac{\tilde{\Omega}(t)}{2}\Delta(t)+iG\Delta(t)\sigma(t) \, ,  \label{A1} \\
\dot{\Delta}(t) &=& i \tilde{\Omega}^{*}(t)\sigma(t)-i \tilde{\Omega}(t)\sigma^{*}(t)+4G_{I}\sigma(t)\sigma^{*}(t) \,. \label{A2}
\end{eqnarray}
Eqs.\ (\ref{A1}) and (\ref{A2}) are the same as the equations of a quantum dot involving only exciton excitation, i.e., modeled by a two-level system, near a metal nanoparticle, see, e.g., Eqs.\ (5) and (6) of Ref.\ \cite{Paspalakis13a}, under exact resonant driving and with no decay and dephasing terms. Therefore, one may use properly the results of Ref.\ \cite{Paspalakis13a} in this case as well. So taking the quantum dot initially in the ground state, we have $\Delta(t=0)=1$ and $\sigma(t=0)=0$, and for a hyperbolic secant pulse, using the result of Eq.\ (11) of Ref.\ \cite{Paspalakis13a}, we directly obtain the result of Eq.\ (\ref{pi2}). Also, for $G=0$, we obtain directly the results of Eq.\ (\ref{pipulse}). In both cases, $\Delta(t) =\sigma_{00}(t)-\sigma_{22}(t) = -\tanh[(t-t_{0})/t_{p}]$, and using Eqs.\ (\ref{r00an})-(\ref{r22an}), which hold even when $E_{B}=0$, we obtain analytical solutions (\ref{anal0}) and (\ref{anal1}).

\section{Analytical treatment using controlled rotation}\label{appenA2}
In this part of the Appendix, we present an analytical treatment of the controlled dynamics of the exciton-biexciton cascade under a single laser excitation using the method of controlled rotation \cite{Paspalakis10a,Kis04b,Paspalakis04b,Kis04a} for $G=0$ and the system at two-photon resonance with the
ground - biexciton transition, i.e., for $\hbar\omega = E + E_{B}/2$. Under these assumptions the Hamiltonian of Eqs.\ (\ref{ham1}) , after a transformation and performing the rotating wave approximation becomes
\begin{equation}
H = -\frac{E_{B}}{2}|1\rangle \langle 1| - \bigg[
\frac{\hbar\bar{\Omega}(t)e^{-i\phi}}{2}|0\rangle \langle 1|
+ \frac{\hbar\bar{\Omega}(t)e^{-i\phi}}{2}|1\rangle \langle 2| + H.c.\bigg] \, , \label{ham3}
\end{equation}
with $\Omega(t) = \bar{\Omega}(t)e^{i\phi}$ and $\bar{\Omega}(t)=|\Omega(t)| = |\Omega_{0}|f(t)$.

The wavefunction of the system can be written in terms of the probability amplitudes as $|\psi(t)\rangle = \sum^{2}_{j=0}b_{j}(t)|j\rangle$. We define
\begin{eqnarray}
b_{0}(t) &=& \frac{e^{-i\phi}}{\sqrt{2}}\left[A(t)-B(t)\right] \, , \\
b_{2}(t) &=& \frac{e^{i\phi}}{\sqrt{2}}\left[A(t)+B(t)\right] \, .
\end{eqnarray}
Using the time-dependent Schr\"{o}dinger equation, the transformed probability amplitudes satisfy the following
equations
\begin{eqnarray}
i\hbar\dot{A}(t) &=& -\frac{\hbar\bar{\Omega}(t)}{\sqrt{2}}b_{1}(t) \, , \label{ampl2a} \\
i\hbar\dot{b}_{1}(t) &=& -\frac{E_{B}}{2} b_{1}(t) - \frac{\hbar\bar{\Omega}(t)}{\sqrt{2}}A(t)
\label{ampl2b} \, , \\
\dot{B}(t) &=& 0 \, .
\end{eqnarray}
Eqs.\ (\ref{ampl2a}) and (\ref{ampl2b}) correspond to a two-level system under the influence of an externally applied field and within the rotating wave approximation. Also, the probability amplitude $B(t)$ does not evolve.
We consider that initially the quantum dot is in its ground state $|0\rangle$, thus $b_{0}(t=0)=1$ and $b_{1}(t=0)=b_{2}(t=0)=0$.

When $b_{1}(t=0) = 0$, the following parametrization may be used for the solutions of the above equations
\begin{equation}
A(t) = \alpha(t) A(0) \, , \quad B(t) = B(0) \, , \quad b_{1}(t) =
\beta(t) A(0) \, ,
\end{equation}
where $\alpha(t), \beta(t)$ satisfy $|\alpha(t)|^2 +
|\beta(t)|^2 = 1$ and are subsequently defined.
Since $A(0) = e^{i\phi}/\sqrt{2}$ and $B(0) = -e^{i\phi}/\sqrt{2}$, for the particular
initial conditions we get
\begin{eqnarray}
b_{0}(t) &=& \frac{1}{2}\left[\alpha(t)+1\right] \, , \label{sol1}\\
b_{1}(t) &=& \frac{e^{i\phi}}{\sqrt{2}}\beta(t) \, , \label{sol2} \\
b_{2}(t) &=& \frac{e^{2i\phi}}{2}\left[\alpha(t) - 1\right]  \, . \label{sol3}
\end{eqnarray}
The occupation probabilities of the three states are given by the analytical expressions
\begin{eqnarray}
\sigma_{00}(t) &=& \frac{1}{4} \left|\alpha (t)+1\right|^{2} \, ,\label{r00an} \\
\sigma_{11}(t) &=& \frac{1}{2}\left|\beta (t)\right|^{2} \, , \label{r11an} \\
\sigma_{22}(t) &=& \frac{1}{4} \left|\alpha (t)-1\right|^{2} \, . \label{r22an}
\end{eqnarray}
From Eqs.\ (\ref{r00an})-(\ref{r22an}) we obtain that perfect
transfer from the ground to the biexciton state is achieved for
$\alpha(t) = -1$ and $\beta(t)=0$.

For zero biexciton energy shift, $E_{B}=0$, and for arbitrary  pulse envelope $f(t)$ [taking that $f(t)=0$ for $t <0$]
\begin{equation}
\alpha(t) = \cos\left[\frac{1}{\sqrt{2}}\int^{t}_{0}\bar{\Omega}(t^{'})dt^{'}\right] \, .
\end{equation}
We choose
\begin{equation} \label{Rescond1}
 |\Omega_{0}|\int^{\infty}_{0}f(t)dt  =  \sqrt{2}\pi \, ,
\end{equation}
which leads to $\alpha(t \rightarrow \infty) = -1$.
In this case, for hyperbolic secant pulse envelope, with $f(t) = \mbox{sech}[(t-t_{0})/t_{p}]$, Eq.\ (\ref{Rescond1}) becomes
\begin{equation}
|\Omega_{0}|= \frac{\sqrt{2}}{t_{p}} \, ,
\end{equation}
and $\alpha(t)$ is
\begin{equation}
\alpha(t) = - \mbox{tanh}\left[\frac{t-t_{0}}{t_{p}}\right] \, .
\end{equation}

For nonzero biexciton shift, $E_{B}\neq 0$, and for hyperbolic secant pulse envelope the Rosen-Zener model
\cite{Rosen32,Vitanov98,Paspalakis10a} is used. Then, $\alpha(t)$ is given by Eq.\ (\ref{hyper}) in the main text. At $t\rightarrow \infty$
\begin{equation}
  \alpha = \frac{\left[\Gamma\left(\frac{1}{2} - i \frac{E_{B}t_{p}}{4\hbar}
      \right)\right]^2} {\Gamma\left(\frac{1}{2} + \frac{|\Omega_{0}|t_{p}}{\sqrt{2}}
      - i \frac{E_{B}t_{p}}{4\hbar}\right)\Gamma\left(\frac{1}{2} - \frac{|\Omega_{0}|t_{p}}{\sqrt{2}} - i
      \frac{E_{B}t_{p}}{4\hbar}\right)} \, ,
\end{equation}
where $\Gamma(\cdot)$ denotes the gamma  function.
A simple formula is found when
\begin{equation}
|\Omega_{0}| =\frac{\sqrt{2}n}{t_{p}}, \quad {\rm with} \quad n = 1,2,\dots \, .
\label{om}
\end{equation}
Then, if we repeatedly apply $\Gamma(z+1) = z \Gamma(z)$
we obtain \cite{Vitanov98}
\begin{eqnarray}
\alpha &=&
(-1)^{n}\prod^{n-1}_{j=0}\frac{j+\frac{1}{2}+i\frac{E_{B}t_{p}}{4\hbar}}{j+
\frac{1}{2}-i\frac{E_{B}t_{p}}{4\hbar}}\, ,\label{rosenzener} \\
\beta &=& 0 \, .
\end{eqnarray}


\end{document}